\documentclass[12pt]{article}
\usepackage{graphicx,epsfig,pstricks,rotating}
\usepackage{bm}

\newcommand{\eqnb}{\begin{equation}}
\newcommand{\eqne}{\end{equation}}

\newcommand{\question}[1]{}
\newcommand{\lsim}[1]{
\setlength{\unitlength}{12pt}
\begin{picture}(1.4,1.)
\put(.7,-0.3){\makebox(0.0,1.)[t]{$<$}}
\put(.7,-0.3){\makebox(0.0,1.)[b]{$\sim$}}
\end{picture}#1}
\begin{document}

\renewcommand{\thefootnote}{\fnsymbol{footnote}}

\begin{center}
{\large \bf Separable Dyson-Schwinger model at finite $T$\footnote[1]{Talk delivered by D. Klabu\v{c}ar}}

\renewcommand{\thefootnote}{\arabic{footnote}}

\vskip 5mm

D. Blaschke$^{\dag,\ddag,}\footnote{{\ttfamily david@theor.jinr.ru}}$,
D. Horvati\' c$^{\star,}\footnote{{\ttfamily davorh@phy.hr}}$,
D. Klabu\v{c}ar$^{\star,}\footnote{{\ttfamily klabucar@oberon.phy.hr}}$,
A. E. Radzhabov$^{\ddag,}\footnote{{\ttfamily aradzh@theor.jinr.ru}}$,

\vskip 5mm

{\small {\it 
$^\dag$ 
Institute for Theoretical Physics, University of Wroclaw, Max Born pl. 9,
50-204 Wroclaw, Poland}}
\\
{\small {\it $^\ddag$ 
Bogoliubov Laboratory of Theoretical Physics,\\ 
Joint Institute for Nuclear Research,
141980 Dubna, Russia}}
\\
{\small {\it $^\star$ Physics Department, Faculty of Science,
University of Zagreb,\\
Bijeni\v{c}ka c. 32, Zagreb 10000, Croatia}}
\end{center}

%$^\star$Talk delivered by D. Klabu\v{c}ar

\begin{abstract}
\noindent
Theoretical understanding of experimental results 
from relativistic heavy-ion collisions requires 
a microscopic approach to the behavior of QCD n-point functions
at finite temperatures, as given by the hierarchy of
Dyson-Schwinger equations, properly generalized within the Matsubara
formalism. The technical complexity of
related finite-temperature calculations however
mandates modeling. We present a model where the 
QCD interaction in the infrared, nonperturbative 
domain, is modeled by a separable form. Results 
for the mass spectrum of light quark flavors %($q=$ $u$, $d$, $s$) 
at finite temperature are presented.
\end{abstract}

\vskip 5mm

\section{Introduction}

While the experiments at RHIC \cite{Muller:2006ee,Adams:2005dq}
advanced the empirical knowledge of the hot QCD matter dramatically, 
the understanding of the state of matter that has been formed is 
still lacking.
For example, the STAR collaboration's assessment \cite{Adams:2005dq} 
of the evidence from RHIC experiments depicts a very intricate, 
difficult-to-understand picture of the hot QCD matter. 
Among the issues pointed out as important was the need
to clarify the role of quark-antiquark ($q\bar q$) bound states
continuing existence above the critical temperature $T_c$, as well
as the role of the chiral phase transition.
%constituent vs. current quarks.

Both of these issues are consistently treated within
the Dyson-Schwinger (DS) approach to quark-hadron physics.
Dynamical chiral symmetry breaking (DChSB) as the crucial low-energy 
QCD phenomenon is well-understood in the rainbow-ladder approximation (RLA),
a symmetry preserving truncation of the hierarchy of DS equations.
Thanks to this, the QCD low energy theorems are fulfilled and the behavior of 
the chiral condensate and pion mass and decay constant are in accord with 
the Gell-Mann--Oakes--Renner relation, i.e. the Goldstone theorem. 

For recent reviews of the DS approach, see, e.g., Refs. 
\cite{Roberts:2000aa,Alkofer:2000wg}, of which the first 
\cite{Roberts:2000aa} also reviews the studies of
QCD  DS equations at finite temperature, started in \cite{Bender:1996bm}.
Unfortunately, the extension of DS calculations to non-vanishing 
temperatures is technically quite difficult. This usage of separable 
model interactions greatly simplifies DS calculations at finite temperatures, 
while yielding equivalent results on a given level of truncation  
\cite{Burden:1996nh,Blaschke:2000gd}. 
A recent update of this covariant separable approach with application to
the scalar $\sigma$ meson at finite temperature can be found in 
\cite{Kalinovsky:2005kx}.
Here, we present results for the quark mass spectrum
at zero and finite temperature, extending previous work by
including the strange flavor.

%%%%%%%%%%%%%%%%%%%%%%%%%%%%%%%%%%

\section{The separable model at zero temperature}
\label{Model}

The dressed quark propagator $S_q(p)$ is the solution of its 
DS equation \cite{Roberts:2000aa,Alkofer:2000wg}, 
%namely the gap equation
\begin{eqnarray}\label{sde}
S_q(p)^{-1} = i \gamma\cdot{p} + \widetilde{m}_q +
\frac{4}{3} \int \frac{d^4\ell}{(2\pi)^4} \, 
g^2 D_{\mu\nu}^{\mathrm{eff}} (p-\ell)
  \gamma_\mu S_q(\ell) \gamma_\nu \, , 
\label{DSE} 
\end{eqnarray}
while the $q\bar q'$ meson Bethe-Salpeter (BS) bound-state vertex 
$\Gamma_{q\bar q'}(p,P)$ is the solution of the BS equation (BSE)
\begin{eqnarray}\label{bse}
-\lambda(P^2)\Gamma_{q\bar q'}(p,P) = \frac{4}{3} \int \frac{d^4\ell}{(2\pi)^4}  
 g^2  D_{\mu\nu}^{\mathrm{eff}} (p-\ell)
\gamma_\mu S_q(\ell_+) \Gamma_{q\bar q'}(\ell,P) S_q(\ell_-) \gamma_\nu \, 
\label{BSE},
\end{eqnarray}
where $D_{\mu\nu}^{\mathrm{eff}}(p-\ell)$ is an effective gluon propagator 
modeling the nonperturbative QCD effects, 
$\widetilde{m}_q$ is the current quark mass, the index $q$ (or $q'$) stands for
the quark flavor ($u, d$ or $s$),
$P$ is the total momentum, and $\ell_{\pm}=\ell\pm P/2$. 
The chiral limit is obtained by setting $\widetilde{m}_q=0$.
The meson mass is identified from $\lambda(P^2=-M^2)=1$.
Equations (\ref{DSE}) and (\ref{BSE}) are written in the Euclidean space,
and in the consistent rainbow-ladder truncation.

The simplest separable Ansatz 
which reproduces in RLA a nonperturbative solution of (\ref{DSE}) for any 
effective gluon propagator in a Feynman-like gauge
$g^2 D_{\mu\nu}^{\mathrm{eff}} (p-\ell) \rightarrow
\delta_{\mu\nu} D(p^2,\ell^2,p\cdot \ell)$ is 
\cite{Burden:1996nh,Blaschke:2000gd}
\begin{eqnarray}
D(p^2,\ell^2,p\cdot \ell)=D_0 {\cal F}_0(p^2) {\cal F}_0(\ell^2) 
+ D_1 {\cal F}_1(p^2) (p\cdot \ell ) {\cal F}_1(\ell^2)~.
\label{sepAnsatz}
\end{eqnarray}
This is a rank-2 separable interaction with two strength parameters $D_i$ 
and corresponding form factors ${\cal F}_i(p^2)$, $i=1,2$. 
The choice for these quantities is constrained to the solution of
the DSE for the quark propagator (\ref{DSE})
\begin{eqnarray}
  S_q(p)^{-1} = i \gamma\cdot{p} A_q(p^2) + B_q(p^2) 
\equiv Z^{-1}_q(p^2) [ i \gamma\cdot{p} + m_q(p^2) ]~,
\end{eqnarray}
where $m_q(p^2)=B_q(p^2)/A_q(p^2)$ is the dynamical mass function and
$Z_q(p^2)=A^{-1}_q(p^2)$ the wave function renormalization.
Using the separable Ansatz (\ref{sepAnsatz}) in (\ref{sde}), 
the gap equations for the quark amplitudes $A_q(p^2)$ and $B_q(p^2)$ read
\begin{eqnarray}
B_q(p^2) - \widetilde{m}_q =  
\frac{16}{3} \int \frac{d^4\ell}{(2\pi)^4} D(p^2,\ell^2,p \cdot \ell)
\frac{B_q(\ell^2)}{\ell^2 A_q^2(\ell^2)+ B_q^2(\ell^2)} 
=  b_q {\cal F}_0(p^2)\, ,
\label{gap1}\\
\left[A_q(p^2)-1 \right]  
=\frac{8}{3p^2} \int \frac{d^4\ell}{(2\pi)^4} D(p^2,\ell^2,p \cdot \ell)
\frac{(p\cdot \ell) A_q(\ell^2)}{\ell^2A_q^2(\ell^2)+B_q^2(\ell^2)} 
= a_q {\cal F}_1(p^2)\, .
\label{gap2}
\end{eqnarray}
Once the coefficients $a_q$ and $b_q$ are
obtained by solving the gap equations (\ref{gap1}) and (\ref{gap2}), 
the only model parameters remaining are $\widetilde{m}_q$ and the parameters 
of the gluon propagator, to be fixed by meson phenomenology.

\section{Extension to finite temperature}

The extension of the separable model studies to the finite temperature case, 
$T\neq 0$, is systematically accomplished by a transcription of the Euclidean 
quark 4-momentum via {$p \rightarrow$} {$ p_n =$} {$(\omega_n, \vec{p})$}, 
where {$\omega_n=(2n+1)\pi T$} are the
discrete Matsubara frequencies. In the Matsubara formalism, the number of
coupled equations represented by (\ref{sde}) and (\ref{bse}) scales up
with the number of fermion Matsubara modes included.  For studies near
and above the transition, \mbox{$T \ge 100 $}~MeV, using only 10 such modes
appears adequate. Nevertheless, the appropriate number can be more than 
$10^3$ if the continuity with \mbox{$T=0$} results is to be
verified. The effective $\bar q q$ interaction will automatically
decrease with increasing $T$ without the introduction of an explicit 
$T$-dependence which would require new parameters.

The solution of the DS equation for the dressed quark propagator 
now takes the form
\begin{eqnarray}
S_q^{-1}(p_n, T) = i\vec{\gamma} \cdot \vec{p}\; A_q(p_n^2,T)
                   + i \gamma_4 \omega_n\; C_q(p_n^2,T)+ B_q(p_n^2,T),\;
\label{invprop}
\end{eqnarray}
where {$p_n^2=\omega_n^2 + \vec{p}^{\,2}$}
and the quark amplitudes 
{$B_q(p_n^2,T) = \widetilde{m}_q + b_q(T) {\cal F}_0(p_n^2)$}, 
\mbox{$A_q(p_n^2,T) = 1+ a_q(T) {\cal F}_1(p_n^2)$}, and 
{$C_q(p_n^2,T) = 1+ c_q(T) {\cal F}_1(p_n^2)$}
are defined by the temperature-dependent coefficients 
$a_q(T)$,$ b_q(T)$, and $c_q(T)$ to be determined from the set of 
three coupled non-linear equations
\begin{eqnarray}
a_q(T) = \frac{8 D_1}{9}\,  T \sum_n \int \frac{d^3p}{(2\pi)^3}\,{\cal F}_1(p_n^2)\, \vec{p}^{\,2}\, [1 + a_q(T) {\cal F}_1(p_n^2)]\; d_q^{-1}(p_n^2,T) \; ,
\\
 c_q(T) = \frac{8 D_1}{3}\,  T \sum_n \int \frac{d^3p}{(2\pi)^3}\,{\cal F}_1(p_n^2)\, \omega_n^2\, [1 +  c_q(T) {\cal F}_1(p_n^2)]\;
                                                d_q^{-1}(p_n^2,T) \; ,
\\
 b_q(T) = \frac{16 D_0}{3}\,  T \sum_n \int \frac{d^3p}{(2\pi)^3}\,{\cal F}_0(p_n^2)\, [\widetilde{m}_q +  b_q(T) {\cal F}_0(p_n^2)]\; d_q^{-1}(p_n^2,T) \; .
\end{eqnarray}
The function $d_q(p_n^2,T)$ is the denominator of the quark 
propagator $S_q(p_n, T)$, and is given by
\begin{eqnarray}
d_q(p_n^2,T) = \vec{p}^{\,2}A_q^2(p_n^2,T) +\omega_n^2C_q^2(p_n^2,T)
                  + B_q^2(p_n^2,T).
\label{Sdenominator}
\end{eqnarray}

The procedure for solving gap equations for a given temperature $T$ is the
same as in $T=0$ case, but one has to control the appropriate
number of Matsubara modes as mentioned above.

\section{Confinement and Dynamical Chiral Symmetry Breaking}

If there are no poles in the quark propagator $S_q(p)$ for real timelike
$p^2$ then there is no physical quark mass shell. This entails 
that quarks cannot propagate freely, and the description of hadronic
processes will not be hindered by unphysical quark production thresholds.
This sufficient condition is a viable possibility for realizing quark 
confinement~\cite{Blaschke:2000gd}. A nontrivial solution for $B_q(p^2)$ 
in the chiral limit (${\widetilde m}_0=0$) signals DChSB.
There is a connection between quark confinement realized as the lack 
of a quark mass shell and the existence of a strong quark mass function 
in the infrared through DChSB.
The propagator is confining if \mbox{$m_q^2(p^2) \neq -p^2$} for real $p^2$,
where the quark mass function is \mbox{$m_q(p^2)=B_q(p^2)/A_q(p^2)$}.
In the present separable model, the strength
\mbox{$b_q=B_q(0)$}, which is generated by solving Eqs. (\ref{gap1})
and (\ref{gap2}), controls both confinement and DChSB.
At finite temperature, the strength $b_q(T)$ for the quark mass function 
will decrease with $T$, until the denominator (\ref{Sdenominator}) of the
quark propagator can vanish for some timelike momentum, and the quark can 
come on the free mass shell. The connection between deconfinement and
disappearance of DChSB is thus clear in the DS approach. Also the
present model is therefore expected to have a deconfinement transition 
at or a little before the chiral restoration transition associated with 
\mbox{$b_0(T) \to 0$}.

The following simple choice of the separable interaction form factors,
$${\cal F}_0(p^2)=\exp(-p^2/\Lambda_0^2)~,~~%$$
%$$
{\cal F}_1(p^2)=\frac{1+\exp(-p_0^2/\Lambda_1^2)}
{1+\exp((p^2-p_0^2)/\Lambda_1^2)}~,$$
is used to obtain numerical solutions which reproduce
very well the phenomenology of the light pseudoscalar mesons and 
generate an acceptable value for the chiral condensate.

The resulting quark propagator is found to be confining and the
infrared strength and shape of quark amplitudes
$A_q(p^2)$ and $B_q(p^2)$ are in quantitative agreement with
the typical DS studies. Gaussian-type form factors are used
as a minimal way to preserve these properties while realizing
that the ultraviolet suppression is much stronger than the asymptotic
fall off (with logarithmic corrections) known from perturbative QCD
and numerical studies on the lattice \cite{Parappilly:2005ei}.

\section{Results}

\begin{figure}[!ht]
\centerline{\includegraphics[width=110mm,angle=0]{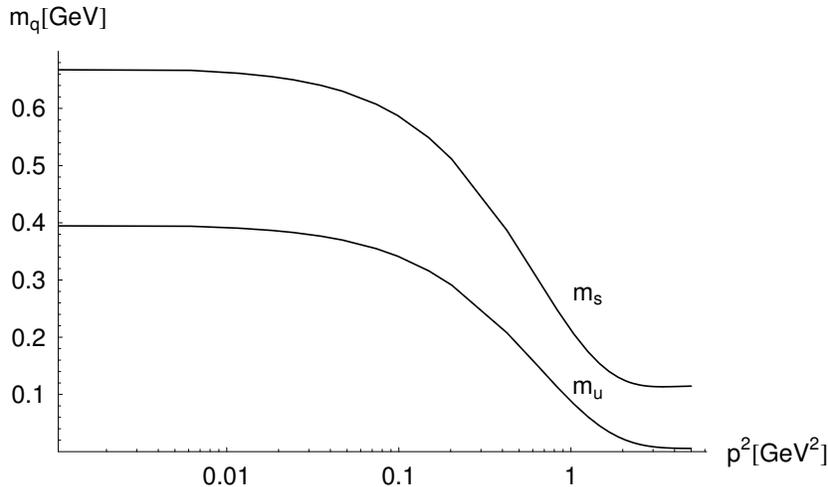}}
\caption{The $p^2$ dependence (at $T=0$) of the dynamically 
generated quark masses $m_s(p^2), m_u(p^2)$ for, respectively,
the strange and the (isosymmetric) nonstrange case.}
\label{figMp2}
\end{figure}

Parameters of the model are completely fixed by meson phenomenology calculated
from the model as discussed in \cite{Blaschke:2000gd,Kalinovsky:2005kx}. 
In the nonstrange sector, we work in the isosymmetric limit and 
adopt bare quark masses ${\widetilde m}_u = {\widetilde m}_d = 5.5$ MeV and
in strange sector we adopt ${\widetilde m}_s = 115$ MeV.
Then the parameter values 
\begin{equation}
\Lambda_0=758 \, {\rm MeV}, \quad 
\Lambda_1=961 \, {\rm MeV}, \quad 
p_0=600 \, {\rm MeV},
\label{Lambda12p0}
\end{equation}
\begin{equation}
D_0\Lambda_0^2=219 \, , \qquad D_1\Lambda_1^4=40 \, ,
\label{D0D1}
\end{equation} 
lead, through the gap equation, to 
$a_{u,d}=0.672$, $b_{u,d}=660$ MeV, $a_{s}=0.657$ and $b_{a}=998$ MeV 
i.e., to the dynamically generated momentum-dependent mass functions
%$m_q(p^2) = B_q(p^2)/A_q(p^2)$ ($q=u,d,s$) 
$m_q(p^2)$ shown in Fig. \ref{figMp2}. 
In the limit of high $p^2$, they converge to ${\widetilde m}_u$ and 
${\widetilde m}_s$.  However, at low $p^2$, the values of 
$m_u(p^2)$ are close to the typical constituent quark 
mass scale $\sim m_\rho/2 \sim m_N/3$ with the maximum value at $p^2=0$, 
$m_u(0)=398$ MeV. 
The corresponding value for the strange quark is $m_s(0)=672$ MeV.
Thus, Fig. \ref{figMp2} shows that in the domain of low
and intermediate $p^2 \lsim 1$ GeV$^2$, the dynamically
generated quark masses have typical constituent
quark mass values. Thus, the DS approach provides a derivation of the 
constituent quark model \cite{Kekez:1998xr} from a more fundamental level, 
with the correct chiral behavior of QCD.

Obtaining such dynamically generated constituent quark masses, 
as previous experience with the DS approach shows (see, e.g.,
Refs. such as
\cite{Roberts:2000aa,Alkofer:2000wg,Kekez:1998xr%,Kekez:2000aw
}),
is essential for reproducing the static
and other low-energy properties of hadrons, including decays.
(We would have to turn to less simplified DS models for incorporating
the correct perturbative behaviors, including that of the quark masses.
Such models are amply reviewed or used in, e.g., Refs.
\cite{Roberts:2000aa,Alkofer:2000wg,Kekez:1998xr,%Kekez:2000aw,
Klabucar:1997zi},
%,Kekez:2001ph,Kekez:2005ie},
but addressing them is beyond the present scope, where the
perturbative regime is not important.)

\begin{figure}[!ht]
\centerline{\includegraphics[width=110mm,angle=0]{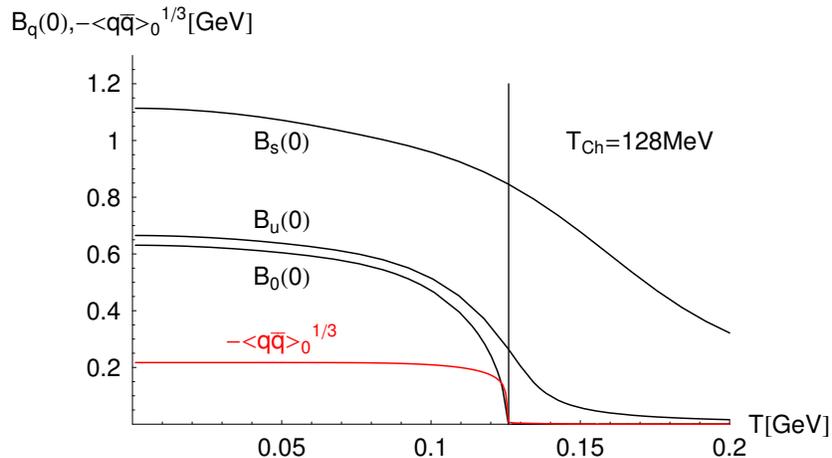}}
\caption{The temperature dependence of $B_s(0), B_u(0)$ and $B_0(0)$, 
the scalar propagator functions at $p^2=0$, for
the strange, the nonstrange and the chiral-limit cases, respectively. 
The temperature dependence of the chiral quark-antiquark condensate,
$-\langle q\bar{q}\rangle^{1/3}_0$, is also shown (by the
lowest curve). Both chiral-limit quantities, 
$B_0(0)$ and $-\langle q\bar{q}\rangle^{1/3}_0$, vanish at the 
chiral-symmetry restoration temperature $T_\mathrm{Ch}=128$ MeV.
}
\label{figBT}
\end{figure}

\begin{figure}[!ht]
\centerline{\includegraphics[width=110mm,angle=0]{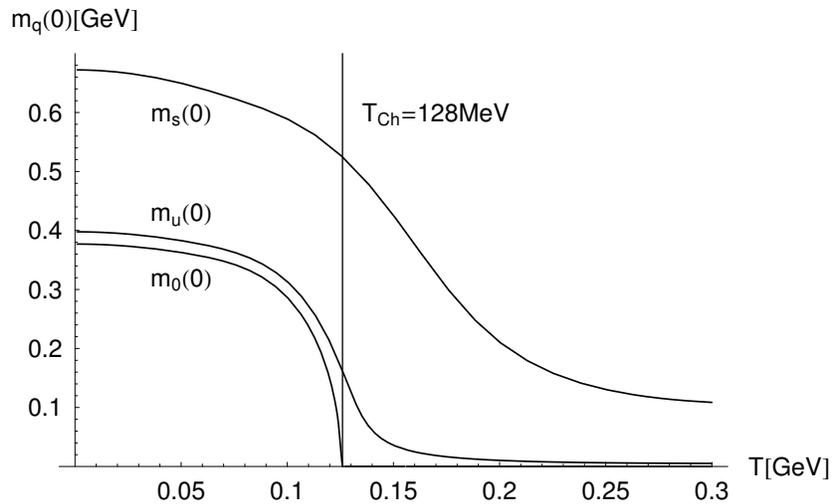}}
\caption{The temperature dependence of $m_s(0), m_u(0)$ and $m_0(0)$, 
the dynamically generated quark masses at $p^2=0$ for 
the strange, the nonstrange and the chiral-limit cases, respectively.}
\label{figMT}
\end{figure}

The extension of results to finite temperatures is given in 
Figs. \ref{figBT}, \ref{figMT}. Very important is the 
temperature dependence of the chiral-limit quantities $B_0(0)_T$ 
and $\langle q \bar q \rangle_{0}(T)$, whose vanishing with $T$ 
determines the chiral restoration temperature $T_\mathrm{Ch}$.
We find $T_\mathrm{Ch} = 128$ MeV in the present model.

\begin{figure}[!ht]
\centerline{\includegraphics[width=110mm,angle=0]{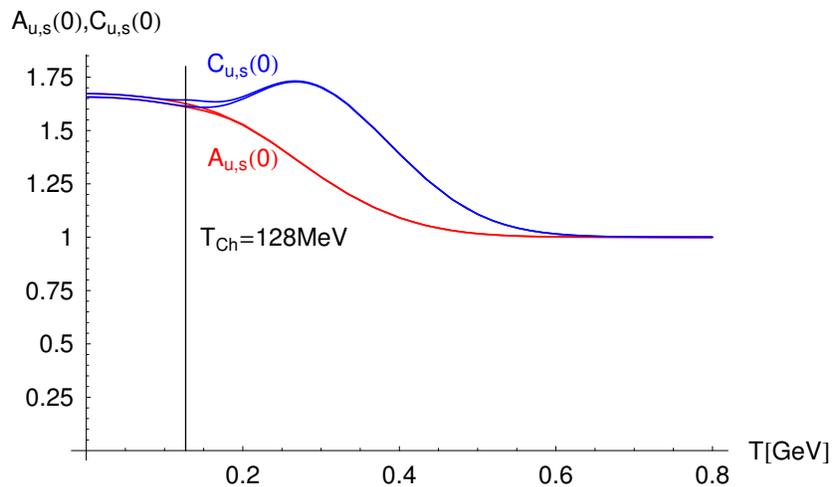}}
\caption{The violation of $O(4)$ symmetry with $T$ is exhibited 
on the example of $A_{u,s}(0)_T$ and $C_{u,s}(0)_T$.}
\label{figACT}
\end{figure}

The temperature dependences of the functions giving the vector part 
of the quark propagator, $A_{u,s}(0)_T$ and $C_{u,s}(0)_T$,
are depicted in Fig. \ref{figACT}. Their difference is a measure
of the O(4) symmetry breaking with the temperature $T$.

The presented model, when applied in the framework of the
Bethe-Salpeter approach to mesons as 
quark-antiquark bound states, produces a very satisfactory 
description of the whole light pseudoscalar nonet, both 
at zero and finite temperatures \cite{Blaschke+alZTF}.

\subsubsection*{Acknowledgments}

We thank M. Bhagwat, Yu.L. Kalinovsky and P.C. Tandy for discussions.
A.E.R.~acknowledges support by RFBR grant No. 05-02-16699, the
Heisenberg-Landau program and the HISS Dubna program of the Helmholtz 
Association. D.H.~and~D.K. were supported by MZT project No.~0119261.
D.B. is grateful for support by the Croatian Ministry of Science for a
series of guest lectures held in the Physics Department at University of
Zagreb, where the present work has been completed. D.K. acknowledges the 
partial support of Abdus Salam ICTP at Trieste, where a part of this 
paper was written.

\end{document}